\documentclass[twoside,a4paper,draft]{article}
\usepackage{latexsym,e-journal}
\usepackage[utf8]{inputenc}
\usepackage[english]{babel}
\usepackage{amssymb,amsfonts}
\usepackage[perpage,symbol*]{footmisc}
\usepackage[final]{graphicx}
\usepackage{pstricks}
\usepackage{cite}
\usepackage[varg]{txfonts}

\begin{document}

\newcommand{\rum}{\rule{0.5pt}{0pt}}
\newcommand{\rub}{\rule{1pt}{0pt}}
\newcommand{\numtimes}{\mbox{\raisebox{1.5pt}{${\scriptscriptstyle \rum\times}$}}}
\newcommand{\numtimess}{\mbox{\raisebox{1.0pt}{${\scriptscriptstyle \rum\times}$}}}
\renewcommand{\refname}{References}
\renewcommand{\tablename}{\small Table}
\renewcommand{\figurename}{\small Fig.}
\renewcommand{\contentsname}{Contents}

\twocolumn[%
\begin{center}
\renewcommand{\baselinestretch}{0.93}
{\Large\bfseries Standard Model Particles from Split Octonions}
\par
\renewcommand{\baselinestretch}{1.0}
\bigskip
Merab Gogberashvili\\
{\footnotesize Javakhishvili Tbilisi State University, 3 Chavchavadze Avenue, Tbilisi 0179, Georgia \\
Andronikashvili Institute of Physics, 6 Tamarashvili Street, Tbilisi 0177, Georgia \rule{0pt}{12pt}\\
E-mail: gogber@gmail.com\\

}\par
\medskip
{\small\parbox{11cm}{%
We model physical signals using elements of the algebra of split octonions over the field of real numbers. Elementary particles are corresponded to the special elements of the algebra that nullify octonionic norms (zero divisors). It is shown that the standard model particle spectrum naturally follows from the classification of the independent primitive zero divisors of split octonions.}}\bigskip\smallskip
\end{center}]{%

\setcounter{section}{0}
\setcounter{equation}{0}
\setcounter{figure}{0}
\setcounter{table}{0}
\setcounter{page}{3}

\markboth{M. Gogberashvili. Standard Model Particles from Split Octonions}{\thepage}
\markright{M. Gogberashvili. Standard Model Particles from Split Octonions}


The algebra of octonions \cite{Sc, Sp-Ve, Baez} is interesting mathematical structure for physical applications (see reviews \cite{Rev-1, Rev-2, Rev-3, Rev-4}). In this paper we suggest that split octonions over the reals form proper mathematical framework to describe elementary particles and show that some physical properties, like the variety of their spices, naturally follows from the structure of the algebra.

In \cite{Gog-FT, Gog, Go-Sa} different aspects of geometrical applications of split octonions over the reals were considered. It is suggested to use split octonions as universal mathematical structure in physics, instead of vectors, tensors, spinors, etc. In this approach world-lines (paths) of particles are parameterized by the elements of split octonions,
\begin{equation} \label{s}
s = \omega + \lambda^nJ_n + x^nj_n + ct I~. ~~~~~ (n = 1, 2, 3)
\end{equation}
Here a pair of repeated upper and lower indices implies a summation, i.e. $x^nj_n = \delta_{nm}x^nj^m$, where $\delta^{nm}$ is Kronecker's delta. 

Four of the eight real parameters in (\ref{s}), $t$ and $x^n$, denote the ordinary space-time coordinates, and $\omega$ and $\lambda^n$ are interpreted as the phase (classical action) and the wavelengths associated with the octonionic signals.

The eight octonionic basis units in (\ref{s}) are represented by one scalar (denoted by $1$), the three vector-like objects $J_n$, the three pseudo vector-like elements $j_n$ and one pseudo scalar-like unit $I$. The squares (inner products) of seven of the hypercomplex basis elements of split octonions give the unit element with the different signs,
\begin{equation} \label{JjI}
J_n^2=1~, ~~~~~ j_n^2=-1~, ~~~~~ I^2=1~.
\end{equation}

It is known that to generate a complete basis of split octonions the multiplication and distribution laws of only three vector-like elements $J_n$ are enough \cite{Sc, Sp-Ve, Baez}. The three pseudo vector-like basis units, $j_n$, in (\ref{s}) can be defined as the binary products,
\begin{equation} \label{jI}
j_n = \frac{1}{2} \varepsilon_{nmk}J^mJ^k~, ~~~~~(n,m,k = 1,2,3)
\end{equation}
where $\varepsilon_{nmk}$ is the totally antisymmetric unit tensor, and thus describe orthogonal planes spanned by vector-like elements $J_n$. The seventh basis unit $I$ (the oriented volume) is defined as the triple product of all three vector-like elements and has three equivalent representation in terms of $J^n$ and $j^n$,
\begin{equation} \label{I}
I = J_1j_1 = J_2j_2 = J_3j_3 ~.
\end{equation}
So the complete algebra of all non-commuting hypercomplex basis units has the form:
\begin{eqnarray} \label{algebra}
J_nJ_m &=& - J_mJ_n = \varepsilon_{nmk} j^k,\nonumber\\
j_nj_m &=& -j_mj_n = \varepsilon_{nmk} j^k,\nonumber\\
j_mJ_n &=& - J_nj_m = \varepsilon_{nmk}J^k, \\
J_nI &=& - IJ_n = j_n~,\nonumber\\
j_nI &=& -Ij_n = J_n~.\nonumber
\end{eqnarray}

The conjugation of vector-like octonionic basis units,
\begin{equation} \label{bar-J}
J_n^\dag = - J_n ~,
\end{equation}
can be understand as reflections. Conjugation reverses the order of $J_n$ in products, i.e.
\begin{eqnarray} \label{bar}
j_n^\dag &=& \frac 12 \left(\varepsilon_{nmk}J^{m}J^{k}\right)^\dag = \frac 12 \varepsilon_{nmk}J^{k\dag}J^{m\dag} = - j_n~, \nonumber \\
I^\dag &=& \left(J_1 J_2 J_3\right)^\dag = J_3^\dag J_2^\dag J_1^\dag= - I~.
\end{eqnarray}
So the conjugation of the pass function (\ref{s}) gives
\begin{equation} \label{s*}
s^\dag = \omega - \lambda_n J^n - x_n j^n - ct I~.
\end{equation}

Using (\ref{JjI}), (\ref{algebra}) and (\ref{s*}) one can find the norm (interval) of the pass function (\ref{s}),
\begin{equation} \label{sN}
N^2 = ss^\dag = s^\dag s = \omega^2 - \lambda^2 + x^2 - c^2t^2~,
\end{equation}
which is assumed to be non-negative. A second condition is that for physical events the vector part of (\ref{s}) should be time-like \cite{Go-Sa},
\begin{equation} \label{time-like}
c^2t^2 + \lambda_n\lambda^n > x_nx^n~.
\end{equation}

One can represent rotations in the space of the split octonions (\ref{s}) by the maps,
\begin{equation} \label{s'}
s' = e^{\epsilon\theta/2 } s e^{-\epsilon\theta/2}~,
\end{equation}
where $\theta$ is some real angle and $\epsilon$ denotes the (3+4)-vector defined by the seven basis units $J_n$, $j_n$ and $I$ \cite{Sc, Sp-Ve, Baez, Go-Sa}. The set of transformations (\ref{s'}), which satisfy the conditions (\ref{sN}) and (\ref{time-like}), form the group $SO(3,4)$ of passive transformations of the coordinates $x^n$, $\lambda^n$ and $t$ \cite{Man-Sch}. However, to represent the active rotations in the space of $s$, which preserves the multiplicative structure (\ref{algebra}) as well, we would need the transformations to be automorphisms. It means not all tensorial transformations of the coordinates $\lambda_n$, $x_n$ and $t$, represent real rotations, only the transformations that have a realization as associative multiplications should be considered. Automorphisms of split-octonions form subgroup of $SO(3,4)$, the noncompact form of Cartan's smallest exceptional Lie group $G_2^{NC}$ \cite{Cart, BHW}.

Infinitesimal transformations of coordinates, which correspond to the action of the main geometrical group of the model, $G_2^{NC}$, can be written as \cite{Go-Sa}:
\begin{eqnarray} \label{x-nu}
x_n' &=& x_n - \varepsilon_{nmk} \alpha^m x^k - \theta_n ct + \nonumber \\
&+&\frac 12 \left( |\varepsilon_{nmk}|\phi^m + \varepsilon_{nmk} \theta^m \right) \lambda^k + \left(\varphi_n - \frac 13 \sum_m \varphi_m\right) \lambda_n~, \nonumber \\
ct' &=& ct - \beta_n \lambda ^n - \theta_nx^n ~, \\
\lambda_n' &=& \lambda_n - \varepsilon_{nmk} \left(\alpha^m - \beta^m\right) \lambda^k + \beta_n ct + \nonumber \\
&+& \frac 12 \left( |\varepsilon_{nmk}| \phi^m - \varepsilon_{nmk} \theta^m \right) x^k + \left(\varphi_n - \frac 13 \sum_m \varphi_m\right) x_n~, \nonumber
\end{eqnarray}
with no summing over $n$ in the last terms of $x_n'$ and $\lambda_n'$. From the 15 parameters (five 3-angles:  $\alpha^m$, $\beta^m$, $\phi^m$, $\theta^m$ and $\varphi^m$) in (\ref{x-nu}), due to the condition
\begin{equation} \label{varphi-1/3}
\sum_n \left(\varphi_n - \frac 13 \sum_m \varphi_m\right) = 0~,
\end{equation}
only 14 are independent.

The transformations (\ref{x-nu}) can be divided into several distinct classes \cite{Go-Sa}. For instance, the $G_2^{NC}$-rotations by the angles $\alpha^n$, $\beta^n$ and $\theta^n$ of the space-time coordinates only, imitate the ordinary infinitesimal Poincar\'{e} transformations of (3+1)-Minkowski space,
\begin{eqnarray} \label{Lorentz}
ct' &=& ct  - \theta_nx^n + a_0 ~, \nonumber \\
x_n' &=& x_n - \varepsilon_{nmk} \alpha^m x^k - \theta_n ct + a_n~,
\end{eqnarray}
where the space-time translations are defined as:
\begin{eqnarray}
a_0 &=& - \beta_n \lambda ^n~, \nonumber \\
a_n &=& \frac 12  \varepsilon_{nmk} \theta^m  \lambda^k~.
\end{eqnarray}
Time translations $a_0$ are smooth, since $\beta_n$ are compact angles, but $\theta^m$ are hyperbolic and for the spatial translations $a_n$ we have the Rindler-like horizons.

Note that Poincar\'{e}-like transformations (\ref{Lorentz}) do not form subgroup of $G_2^{NC}$ (the subgroup structure of $G_2^{NC}$ one can be find, for example, in \cite{BHW}), since we had neglected rotations of the extra time-like parameters $\lambda_n$. Complete $G_2^{NC}$-transformations reveal some new features in compare to the Minkowski case, like parity violations \cite{Go-Sa}.

Another class of automorphisms,
\begin{eqnarray} \label{varphi}
x_n' &=& x_n + \left(\varphi_n - \frac 13 \sum_m \varphi_m \right) \lambda_n~, \nonumber \\
t' &=& t ~, \\
\lambda_n' &=& \lambda_n  + \left(\varphi_n - \frac 13 \sum_m \varphi_m \right) x_n~, \nonumber
\end{eqnarray}
represent rotations through hyperbolic angles, $\varphi_1$, $\varphi_2$ and $\varphi_3$ (of the three, due to (\ref{varphi-1/3}), only two are independent) of the pairs of space-like and time-like coordinates $x_n$ and $\lambda_n$, into the orthogonal planes $(x_1 - \lambda_1)$, $(x_2 - \lambda_2)$ and $(x_3 - \lambda_3)$. It is convenient to define 2-parameter Abelian subalgebra of $G_2^{NC}$ by the generators of two independent rotations in these planes. It is known that the rank of $G_2^{NC}$ is two, as of the group $SU(3)$ \cite{BHW, Gun-Gur}. In terms of the two parameters, $K_1$ and $K_2$, which are related to the angles $\varphi_n$ as
\begin{eqnarray} \label{K}
K_1 &=& \frac 13 \left(2\varphi_1 - \varphi_2 - \varphi_3\right)~, \nonumber \\
K_2 &=& -  \frac {1}{2\sqrt 3} \left(2 \varphi_3 - \varphi_1 - \varphi_2 \right)~,
\end{eqnarray}
the transformations (\ref{varphi}) can be written more concisely,
\begin{equation} \label{l-x}
\begin{pmatrix}
{\lambda_1' + I x_1'\cr
\lambda_2' + I x_2'\cr
\lambda_3' + I x_3'} 
\end{pmatrix}
= e^{\left(K_1\Lambda_3 + K_2 \Lambda_8\right) I}
\begin{pmatrix}
{\lambda_1 + I x_1 \cr
\lambda_2 + I x_2 \cr
\lambda_3 + I x_3} 
\end{pmatrix}~,
\end{equation}
where $I$ is the vector-like octonionic basis unit ($I^2 = 1$) and $\Lambda_3$ and $\Lambda_8$ are the standard $3\times 3$ Gell-Mann matrices \cite{Go-Sa}. Then one can classify irreducible representations of $G_2^{NC}$ by two fundamental simple roots of the algebra ($K_1$ and $K_2$) and using analogies with $SU(3)$ interpret them as corresponding to the spin and hypercharge of particles. It is known that all quarks, antiquarks, and mesons can be imbedded in the adjoint representation of $G_2^{NC}$ \cite{Gun-Gur}.

In the approach \cite{Gog-FT, Gog, Go-Sa} the norm (\ref{sN}) can be viewed as some kind of space-time interval with four time-like dimensions. The ordinary time parameter, $t$, corresponds to the distinguished octonionic basis unit, $I$, while the other three time-like parameters, $\lambda_n$, have a natural interpretation as wavelengths, i.e. do not relate to time as conventionally understood. Within this picture, in front of time-like coordinates in the expression of pseudo-Euclidean octonionic intervals there naturally appear two fundamental physical parameters, the light speed and Planck's constant. Then from the requirement of positive definiteness of norms under $G_2^{NC}$-transformations, together with the introduction of the maximal velocity, there follow conditions which are equivalent to uncertainty relations \cite{Gog, Go-Sa}. Also it is known that a unique physical system in multi-time formalism generates a large variety of 'shadows' (different dynamical systems) in (3+1)-subspace \cite{times-1, times-2, times-3, times-4, times-5}. One can speculate that  information of multi-dimensional structures, which is retained by these images of the initial system, might takes the form of hidden symmetries in the octonionic particle Lagrangians \cite{Go-Sa}.

Split algebras contain special elements with zero norms (zero divisors) \cite{Sc}, which are important structures in physical applications \cite{So}. For the coordinate function (\ref{s}) we can define the deferential zero divisor,
\begin{equation} \label{d}
\frac {d}{ds} = \frac 12 \left[ \frac {d}{d\omega} - J_n \frac {d}{d\lambda_n} - j_n \frac {d}{dx_n} - I\frac {d}{cdt}\right]~,
\end{equation}
such that its action upon $s$ is:
\begin{equation}
\frac {ds}{ds} = 1~.
\end{equation}
The operator (\ref{d}) annihilates $s^\dag$, while the conjugated derivative operator,
\begin{equation} \label{d*}
\frac {d}{ds^\dag} = \frac 12 \left[ \frac {d}{d\omega} + J_n \frac {d}{d\lambda_n} + j_n \frac {d}{dx_n} + I\frac {d}{cdt}\right]~,
\end{equation}
is zero divisor for $s$, i.e.
\begin{equation}
\frac {ds^\dag}{ds} = \frac {ds}{ds^\dag} = 0~.
\end{equation}
From these relations it is clear that the interval (\ref{sN}) is a constant function for the restricted left octonionic gradient operators,
\begin{eqnarray}
\frac {d}{ds_L} \left(s^\dag s\right) = \left(\frac {ds^\dag}{ds} \right) s = 0~, \nonumber \\
\frac {d}{ds^\dag_L} \left(ss^\dag\right) = \left(\frac {ds}{ds^\dag} \right) s^\dag = 0~,
\end{eqnarray}
and the invariance of the intervals,
\begin{equation}
ds^2 = ds ds^\dag = ds^\dag ds~,
\end{equation}
in the space of split octonions can be viewed as an algebraic property.

The octonionic wavefunctions $\Psi$, in general, should depend on $s$ and on $s^\dag$ as well. Thus we need the condition of analyticity of functions of split octonionic variables,
\begin{equation} \label{dPsi/ds}
\frac {d\Psi(s,s^\dag)}{ds^\dag} = 0~,
\end{equation}
which is similar to the Cauchy-Riemann equations from complex analysis. It can be shown that the system of eight algebraic conditions (\ref{dPsi/ds}), in certain cases \cite{Gog-Split}, lead to the octonionic Maxwell and Dirac equations \cite{Gog-FT}.

Now consider non-differential zero divisors. These type of quantities are distinct elements of the algebra and thus in physical applications could be corresponded to the unit signals (elementary particles). In the algebra of split octonions two types of primitive zero divisors, idempotent elements (projection operators) and nilpotent elements (Grassmann numbers), can be constructed \cite{Sc, Go-Sa}. There exist four classes (totally eight) of primitive idempotents,
\begin{eqnarray} \label{D}
D^{\pm}_n &=& \frac 12 \left(1 \pm J_n \right)~, ~~~~~~~~~~~(n = 1,2,3) \nonumber \\
d^{\pm} &=& \frac 12 \left(1 \pm I\right) ~,
\end{eqnarray}
which obey the relations:
\begin{eqnarray} \label{DD}
D^{\pm}_n D^{\pm}_n &=& D^{\pm}_n~, \nonumber \\
d^{\pm}d^{\pm} &=& d^{\pm} ~.
\end{eqnarray}
The pairs $(D^+_n, D^-_n)$ and $(d^+, d^-)$ are zero divisors for each other,
\begin{eqnarray}
D^{\pm}_n D^{\mp}_n &=& 0 ~, \nonumber \\
d^{\pm}d^{\mp} &=& 0~,
\end{eqnarray}
and thus commute,
\begin{equation} \label{[DD=0]}
[D^+_n, D^-_n] = [d^+, d^-] = 0~.
\end{equation}

We have also twelve classes (twenty four in total) of primitive nilpotents,
\begin{eqnarray} \label{G}
G^{\pm}_{nm} &=& \frac 12 \left(J_n \pm j_m\right) ~, ~~~~~~~~~~~ (n, m = 1,2,3) \nonumber \\
g^{\pm}_n &=& \frac 12 \left(I \pm j_n \right) ~,
\end{eqnarray}
which are zero divisors for themselves,
\begin{eqnarray} \label{GG}
G^{\pm}_{nm} G^{\pm}_{nm} &=& 0~, \nonumber \\
g^{\pm}_n g^{\pm}_n &=& 0 ~.
\end{eqnarray}
We see that separately the quantities (\ref{G}) can be considered as the Grassmann numbers, but do not commute with their conjugates,
\begin{eqnarray} \label{G+G-}
G^\pm_{nn} G^\mp_{nn} &=& d^\mp ~, \nonumber \\
G^\pm_{nm} G^\mp_{nm} &=& \epsilon_{nmk}D^\pm_k~, ~~~~~ n\ne m ~~ (n,m,k = 1,2,3) \\
g^\pm_n g^\mp_n &=& D^\pm_n ~, \nonumber
\end{eqnarray}
in contrast to the case of projection operators (\ref{[DD=0]}). The quantities $G^\pm_{nm}$ and $g^\pm_n$ are the elements of so-called algebra of Fermi operators with the anti-commutators,
\begin{eqnarray}
G^\pm_{nm}G^\mp_{nm} + G^\mp_{nm}G^\pm_{nm} &=& 1~, \nonumber \\
g^\pm_ng^\mp_n + g^\mp_ng^\pm_n &=& 1~,
\end{eqnarray}
which is some syntheses of the Grassmann and Clifford algebras.

We want to emphasize that the number of distinct primitive idempotents (four) and nilpotents (twelve), and there conjugates, coincides with the number of particle/antiparticle spices (bosons and fermions, respectively) of the standard model. This justifies our assumption that primitive zero divisors, which describe unit signals in the space of split octonions, can be corresponded to the elementary particles. The properties that the product of two projection operators reduces to the same idempotent (\ref{DD}), while the product of two Grassmann numbers is zero (\ref{GG}), naturally explains the validity of the Bose and Fermi statistics for the corresponding particles. In this picture distinct statistics follows from the existence of the two types of 'light-cones' in the octonionic (4+4)-space (\ref{sN}), what shows itself in the definitions of the primitive zero divisors (\ref{D}) and (\ref{G}). Also note that the number of the standard model particle generations and the amount of spatial dimensions, both follow from the structure of the algebra of split octonions and are connected with the exitance of the three fundamental vector-like elements $J_n$.

To conclude, in this paper geometrical applications of real split octonions are considered and elementary particles are connected with zero divisors, the special elements of the algebra which nullify octonionic intervals. It is shown that the standard model particle spectrum naturally follows from the classification of the independent primitive zero divisors of the algebra.

\vskip 3mm
\noindent \footnotesize
{\bf Acknowledgments:} This research was supported by the Shota Rustaveli National Science Foundation grant ${\rm ST}09\_798\_4-100$.


\vspace*{-6pt}
\centerline{\rule{72pt}{0.4pt}}
}


\begin{thebibliography}{99}\footnotesize

\bibitem{Sc} Schafer R.,
            \textit{Introduction to Non-Associative Algebras}, Dover, NY 1995.

\bibitem{Sp-Ve} Springer T.A and Veldkamp F.D.
               \textit{Octonions, Jordan Algebras and Exceptional Groups}, Springer Monographs in Mathematics,
               Springer, Berlin 2000.
               
\bibitem{Baez} Baez J.C. \textit{Bull. Am. Math. Soc.}, 2002, v.39, 145, arXiv: math/0105155 [math.RA].

\bibitem{Rev-1} Okubo S.,
               \textit{Introduction to Octonion and Other Non-Associative Algebras in Physics},
               Cambrodge Univ. Press, Cambridge 1995.

\bibitem{Rev-2} Finkelstein D.,
               \textit{Quantum Relativity: A Synthesis of the Ideas of Einstein and Heisenberg},
               Springer, Berlin 1996.

\bibitem{Rev-3} G\"{u}rsey F. \& Tze C.
               \textit{On the Role of Division, Jordan and Related Algebras in Particle Physics},
               World Scientific, Singapore 1996.

\bibitem{Rev-4} L\~{o}hmus J., Paal P. \& Sorgsepp L.
               \textit{Nonassociative Algebras in Physics},
               Hadronic Press, Palm Harbor 1994;
               \textit{Acta Appl. Math.}, 1998, v. 50, 3.

\bibitem{Gog-FT} Gogberashvili M.
                \textit{Int. J. Mod. Phys.}, 2006, v. A 21, 3513, arXiv: hep-th/0505101;
                \textit{J. Phys.}, 2006, v.  A 39, 7099, arXiv: hep-th/0512258.

\bibitem{Gog} Gogberashvili M.
             arXiv: hep-th/0212251;
             \textit{Adv. Appl. Clif. Alg.}, 2005, v. 15, 55, arXiv: hep-th/0409173;
             \textit{Adv. Math. Phys.}, 2009, v. 2009, 483079, arXiv: 0808.2496 [math-ph].

\bibitem{Go-Sa} Gogberashvili M. \& Sakhelashvili O.
               \textit{Adv. Math. Phys.}, 2015, v. 2015, 196708, arXiv: 1506.01012 [math-ph].

\bibitem{Man-Sch} Manogue C.A. \& Schray J.
                 \textit{J. Math. Phys.}, 1993, v. 34, 3746.

\bibitem{Cart} Cartan E.
              \textit{Sur la structure des groupes de transformations finis et continus},
              These, Paris 1894, p. 146;
              \textit{Ann. Sci. Ecole Nonn. Sup.}, 1914, v. 31, 263;
              Reprinted in: \textit{Oeuvres compl\`{e}tes}, Gauthier-Yillars, Paris 1952.

\bibitem{BHW} Beckers J., Hussin V. \& Winternitz P.
             \textit{J. Math. Phys.}, 1986, v. 27, 2217.

\bibitem{Gun-Gur} G\"{u}naydin M. \& G\"{u}rsey F.
                 \textit{J. Math. Phys.}, 1973, v. 14, 1651;
                 \textit{Lett. Nuovo Cim.}, 1973, v. 6, 401;
                 \textit{Phys. Rev.}, 1974, v. D 9, 3387.

\bibitem{times-1} Cole E.A.B. 
                 \textit{Nuov. Cim.}, 1977, v. A 40, 171;
                 \textit{J. Phys.}, 1980, v. A 13, 109.

\bibitem{times-2} Gogberashvili M.
                 \textit{Phys. Lett.}, 2000, v. B 484, 124, arXiv: hep-ph/0001109.

\bibitem{times-3} Gogberashvili M. \& Midodashvili P.
                \textit{Phys. Lett.}, 2001, v. B 515, 447, arXiv: hep-ph/0005298;
                \textit{Europhys. Lett.}, 2003, v. 61, 308, arXiv: hep-th/0111132.

\bibitem{times-4} Christian J.
                 \textit{Int. J. Mod. Phys.}, 2004, v. D 13, 1037, arXive: gr-qc/0308028;
                 in \textit{Relativity and the Dimensionality of the World}, ed. V. Petkov,
                 Springer, NY 2007, arXiv: gr-qc/0610049.

\bibitem{times-5} Velev M.V.
                 \textit{Phys. Essays}, 2012, v. 25, 3.

\bibitem{So} Sommerfeld A.
            \textit{Atombau und Spektrallinien}, II Band,
            Vieweg, Braunschweig 1953.

\bibitem{Gog-Split} Gogberashvili M.
                   \textit{Eur. Phys. J.}, 2014, v. C 74, 3200, arXiv: 1410.4136 [physics.gen-ph].

\end{thebibliography}
\end{document}